# Hydrogenation *via* a low energy mechanochemical approach: the MgB$_2$ case

Claudio Pistidda,[a] Archa Santhosh,[a] Paul Jerabek,[a] Yuanyuan Shang,[a] Alessandro Girella,[b] Chiara Milanese,[b] Maria Dore,[c] Sebastiano Garroni,[c] Simone Bordignon,[d] Michele R. Chierotti,[d] Thomas Klassen,[a,e] Martin Dornheim[a]

[a]Institute of Hydrogen Technology, Materials Design, Helmholtz-Zentrum Geesthacht GmbH, Max-Planck Strasse 1, Geesthacht, D-21502, Germany

[b]Pavia Hydrogen Lab, Chemistry Department, Physical Chemistry Section, C.S.G.I. and Pavia University, Viale Taramelli, 16, Pavia, 27100, Italy

[c]Università degli Studi di Sassari, Dipartimento di Chimica e Farmacia, via Vienna 2, Sassari, 07100, Italy

[d]Department of Chemistry and NIS Centre, University of Torino, Via Giuria 7, Torino, 10125, Italy

[e]Institute of Materials Technology, Helmut-Schmidt-University, Hamburg, 22043, Germany

Corresponding author: Claudio Pistidda claudio.pistidda@hzg.de

**Abstract**

This work aims at investigating the effect that the energy transferred during powder-to-wall collisions in a milling process without grinding media entails on solid-gas reactions. For this purpose, the synthesis of Mg(BH$_4$)$_2$ from MgB$_2$ in a pressurized hydrogen atmosphere was chosen as a model reaction. MgB$_2$ was milled under a broad set of milling parameters (*i.e.* milling times and rotation regimes) and the obtained product thoroughly characterized. By proving the partial formation of Mg(BH$_4$)$_2$, the results of this investigation indicate that the energy transferred to the powder bed by the powder particles colliding with the chamber wall during milling is not negligible, in particular when the milling process is protracted for a long period.

**Introduction**

Mechanochemistry is the branch of chemistry that studies processes that occur under mechanical energy input. Since the 1860s, the development of mechanochemical techniques increasingly influenced several technological fields among which is powder processing. These techniques found application not only in the production of fine and homogeneous powder mixtures, but also for the synthesis of advanced materials.[1]

Mechanochemical processes are usually carried out *via* the use of high-energy mills *e.g.* planetary mills, roto-vibrational mills, vibration mills, attritor mills, pin mills, and rolling mills.

The milling process takes place into a specifically designed milling chamber. Several different variables influence the milling outcome, *e.g.* type of the mill, material of the grinding media, ball-to-powder ratio (BTP), filling extent of the milling chamber, milling atmosphere, milling speed, milling time, etc...[2]

Upon milling, in milling devices such as planetary mills, mechanical stresses are applied on a fraction of powder particles trapped between grinding media colliding with the wall of the milling chamber. When the mechanically applied stresses overcome the material yield point of the trapped material, complex sequences of local atomic rearrangement and subsequent equilibration stages lead to phenomena of cold work, fracture, and plastic deformation of the material particles. This whole sequence of events can be regarded as a modification of the atomic coordination shells, which locally generates a transient energy excess in the crystal lattice. The region of the solid where the energy excess is located moves away from equilibrium conditions generating a so-called local excited state (LES).[3] The formation of LESs is considered to be the reason for the unusual chemical behavior observed in processes taking place under mechanochemical input (*e.g.* enhanced absorption of gaseous phases). In a milling process, the energy transferred as the consequence of collisions between powder particles and chamber wall is not considered. This is mostly due to the expected low relevance that the energy transferred through these collisions has on the overall mechanochemical process when compared to the energy transferred during the impacts of the grinding media with the chamber walls.

In the past decades, mechanochemical approaches have been extensively utilized in the synthesis of several hydrogen storage materials.[4-23] In addition, it is expected that mechanochemical methods hold significant unexplored potential for the development of novel materials, *e.g.* for 'green chemistry'.[24]

Among the hydrogen storage materials that can be synthesized *via* mechanochemical methods, metal borohydrides, owing to their large hydrogen storage capacities, are considered as potential hydrogen storage candidates for mobile and stationary applications.[25-31]

In particular, magnesium borohydride (*i.e.* $Mg(BH_4)_2$), owing to a gravimetric hydrogen capacity of 14.9%, a volumetric hydrogen capacity of 147 kg/m$^3$, and an evaluated enthalpy of decomposition of about 40 kJ/mol$_{H2}$, is considered as one of the most important boron-based hydrogen storage compounds.[32-34,35,36] The uncertainty about the $Mg(BH_4)_2$ enthalpy of decomposition is due to the fact that, depending on the applied experimental conditions, $Mg(BH_4)_2$ might decompose following different and or multiple decomposition reaction paths, as described in equations 1-4.

$$Mg(BH_4)_2 \rightarrow Mg + 2B + 4H_2 \quad (1)$$

$$Mg(BH_4)_2 \rightarrow MgH_2 + 2B + 3H_2 \quad (2)$$

$$Mg(BH_4)_2 \rightarrow 1/6MgB_{12}H_{12} + 5/6MgH_2 + 13/6H_2 \rightarrow MgB_2 + 4H_2 \quad (3)$$

$$Mg(BH_4)_2 \rightarrow MgB_2 + 4H_2 \quad (4)$$

However, at high temperatures (*i.e.* above 500 °C) the final solid decomposition product of $Mg(BH_4)_2$ is $MgB_2$. It must be noticed that during the decomposition process the possibility of releasing small quantities of $B_2H_6$ was also reported.[5,32,35,37] In literature four different crystalline magnesium borohydride polymorphs are reported, *i.e.* hexagonal α (P6$_1$22),[33,38-40] orthorhombic β (Fddd),[35,41] cubic γ (Id-3a)[42] and trigonal ζ (P-3m1)[43]. Depending on the utilized synthesis method, it is possible to obtain magnesium borohydride also in an amorphous state.[44] The synthesis of $Mg(BH_4)_2$ *via* wet chemistry methods is challenging since several possible solvent adducts can be formed (*e.g.* diethyl ether, toluene/heptane, and amine solutions).[38] Solvent-free processes to synthesize magnesium borohydride based on the metathesis reaction of $MgCl_2$ with $NaBH_4$ and $LiBH_4$ performed under mechanochemical input were also reported.[32] The possibility to convert $MgB_2$ into β-$Mg(BH_4)_2$ with a conversion yield of 75% was reported to be possible at 400 °C and under a hydrogen pressure of 950 bar.[37] According to the cited literature, the hydrogenation pathway of $MgB_2$ is a multi-step process occurring at molecular scale in which the

$H_2$ dissociation at the $MgB_2$ surface is followed by the migration of atomic hydrogen to boron sites, where the formation of stable B–H bonds leads to the generation of $Mg(BH_4)_2$. Recently, Pistidda et al. reported for the first time on the possibility to form amorphous $Mg(BH_4)_2$ through high-energy ball milling (BM) of $MgB_2$ under an atmosphere of hydrogen pressurized to 100 bar. In this work, a conversion yield of $MgB_2$ into amorphous $Mg(BH_4)_2$ of roughly 50% was achieved after 100 hours of milling under a starting $p_{H2}$ of 100 bar using a ball-to-powder ratio (BTP) of 30:1 and a rotation speed of 600 rpm.[5]

In this work, the synthesis of $Mg(BH_4)_2$ *via* the hydrogenation of $MgB_2$ is used as a model reaction to investigate the effect that the energy transferred during powder-to-wall collisions in a milling process might entail. For this reason, several different batches of $MgB_2$ were charged into a milling chamber which was subsequently pressurized with hydrogen at 100 bar and milled (without the use of grinding media) for increasing times under a constant rotation regime (rpm) or under increasing rotation regimes for a fixed time.

**Experimental**

The $MgB_2$ used in this work was purchased from Alfa Aesar in powder form with a purity equal to 95%. The material were mixed (no grinding media were used) under a reactive atmosphere of hydrogen by using a stainless steel pressure chamber from Evico Magnetics mounted on a Fritsch Planetary Mono Mill PULVERISETTE 6. Two $MgB_2$ batches of 15 g were used. The first batch was the as-received $MgB_2$: this material was divided into three parts (5 grams for each test), which were then mixed for 10, 25, and 50 hours at 550 rpm, respectively, under 100 bar of hydrogen. The second batch was prepared starting from the pre-milled $MgB_2$ (10 hours at 550 rpm using a BTP ratio of 10:1) divided into three parts (5 grams for each test) and then mixed for 50 hours at 350, 450, and 550 rpm, respectively, under 100 bar of hydrogen. For the sake of clarity, the samples are named in the following way: commercial $MgB_2$ is reported as As received and the same material after 10, 25, and 50 hours of mixing at 100-bar $p_{H2}$ are reported as 10 h, 25 h, and 50 h respectively. Instead, the as-received $MgB_2$ milled for 10 hours in a Mono Mill PULVERISETTE 6 is reported as Milled and the same material after 50 hours of mixing under 100-bar $p_{H2}$ at 350, 450, and 550 rpm are reported as 350 rpm, 450 rpm, and 550 rpm, respectively. Powder handling and milling were carried out in dedicated glove boxes under a continuously purified argon flow (< 2 ppm of $O_2$ and $H_2O$). *Ex-situ* powder X-ray diffraction analyses (PXD) were performed in-house using a Siemens D5000 X-ray diffractometer equipped with a Cu source using the Kα radiation (λ = 1.54056 Å) in the Bragg–Brentano geometry. Inside an argon filled glove box (< 1 ppm of $O_2$ and $H_2O$), the powder was dispersed onto a silicon single crystal sample holder, airtight-sealed with a Kapton film. The microstructural parameters were investigated by fitting the diffraction patterns with the program MAUD (Materials Analysis Using Diffraction) using the Rietveld method.[45] The material thermal stability was investigated in the range of temperature between 20 and 500 °C *via* differential thermal analysis (DTA) using a NetzschSTA409 machine. The DTA analyses were performed using a constant argon flow of 50 mL/min and heating rates of 1, 3, and 5 °C/min in open $Al_2O_3$ crucibles. For each DTA analysis, about 30 mg of material was charged in the crucibles. The evolution of $H_2$ and $B_2H_6$ from the samples heated in the DTA equipment at a heating rate of 3 °C/min were monitored using a HPR20 Benchtop Gas Analysis System from Hiden Analytical.

The volumetric analysis of all the investigated samples was performed using a Sievert´s type apparatus (Hera, Quebec, Canada). The measurements were performed at a $p_{H2}$ of 1 bar, heating the material from about 90 °C up to 450 °C (heating rate of 3 °C/min) and then keeping it isothermally at 450 °C for several minutes. The material morphology was characterized by scanning electron microscopy (SEM), using an EvoMA10 microscope (Zeiss, Germany) equipped with a $LaB_6$ filament. To avoid possible oxygen and moisture contaminations, an in-house built sample holder was used. The samples were charged into the sample holder in an argon-filled glove box and then transported to the SEM.

The composition of the gaseous phase released during the desorption reaction of the investigated samples was analyzed using a Hiden Analytical HAL 201 Mass-Spectrometer, which is coupled with a Netzsch STA 409 C Differential Thermal Analysis (DTA-MS). About 5 mg of each sample was placed in a $Al_2O_3$ crucible that was heated from room temperature up to 500 °C and then cooled down to room temperature in the DTA apparatus, with a heating rate of 3 °C/min. The measurements were performed under a continuous Ar flow of 50 mL/min.

The Solid-State Nuclear Magnetic Resonance (SSNMR) investigations were carried out using a Bruker Avance or Avance II 400 MHz spectrometers operating at 400.2 and 128.3 MHz for $^1H$ and $^{11}B$ nuclei, respectively. The investigated samples were packed in a glove box under Ar atmosphere into cylindrical zirconia rotors with a 4 mm o.d. and an 80 µL volume. The $^{11}B$ direct excitation magic-angle spinning (MAS NMR) spectra of Figure 5 were acquired using a spinning speed of 12 kHz in a dry nitrogen gas flow. The $^1H$-$^{11}B$ cross-polarization (CPMAS) spectrum of the 550 rpm sample (shown in Figure 6) was acquired at room temperature at a spinning speed of 10 kHz, using a ramp cross-polarization pulse sequence with a 90° $^1H$ pulse of 3.80 µs, a contact time of 1 ms, an optimized recycle delay of 3.2 s and a number of scans of 32750. A two-pulse phase modulation (TPPM) decoupling scheme was used, with a radiofrequency field of 69.4 kHz. The $^{11}$B chemical shift scale was calibrated through the $^{11}$B signal of external standard NaBH$_4$ (at -42.0 ppm with respect to BF$_3$·O(CH$_2$CH$_3$)$_2$). The spinning sidebands are indicated by the asterisk symbol "*".

The first-principles calculation of NMR parameters was carried out with the gauge-including projector augmented-wave (GIPAW)[46] and the projector augmented-wave (PAW) method as implemented in the Vienna *ab initio* simulation package (VASP)[47]. Exchange and correlation effects were described using the generalized gradient approximation (GGA) with the Perdew–Burke-Ernzerhof (PBE) functional.[48] The plane wave expansion includes all plane waves within a kinetic energy cut-off of 600 eV. The sampling of the Brillouin zone was performed using a Monkhorst-Pack scheme with a dense k-mesh corresponding to a k-spacing of 0.2 Å. The structures were optimized prior to GIPAW calculations until the forces on each atom were less than 0.01 eV/Å.

GIPAW calculations yielded the principal components of the absolute shielding tensor (σ). The isotropic chemical shift ($δ_{iso}$) could then be determined by comparing a model system computed at the same level to the known experimental shift ($δ_{iso}^e$) to obtain the reference isotropic shielding ($δ_{iso}^r$). In this study, $δ_{iso}^r$ is set as -101.39 ppm to fit the $δ_{iso}^e$ of Mg(BH$_4$)$_2$ and MgB$_2$.

**Results**

The starting materials, and the same after mixing in hydrogen atmosphere, were characterized by means of *ex-situ* PXD. The acquired diffraction patterns are displayed in Figure 1. The patterns of As received and of the same material after high-frequency mixing for increasing time under hydrogen atmosphere (10 h, 25 h, and 50 h) are shown on the left-hand side (A). The patterns of the ball-milled material (Milled) and of the same one after milling (without grinding media) in a hydrogen atmosphere under different rotation regimes (350 rpm, 450 rpm and 550 rpm) are shown on the right-hand side (B). All the specimens appear to be composed mostly of crystalline MgB$_2$ plus a small amount of magnesium. Despite the apparently identical composition, for the MgB$_2$ phase it is possible to perceive a progressive increment of the value of full width at half maximum (FWHM) of the material treated under a hydrogen atmosphere. This effect is more marked for the material that underwent ball milling prior to the hydrogen treatment (Fig 1B). Considering that the diffraction measurements were acquired all at the same temperature (room temperature) and using the same instrumental setup, the broadness of diffraction peaks can be related only to the size of the MgB$_2$ crystallites and to its lattice strains. In order to quantify variations of the crystallite size and lattice strain, the Rietveld refinement of all the diffraction patterns was carried out. The refinement results are summarized in Table 1.

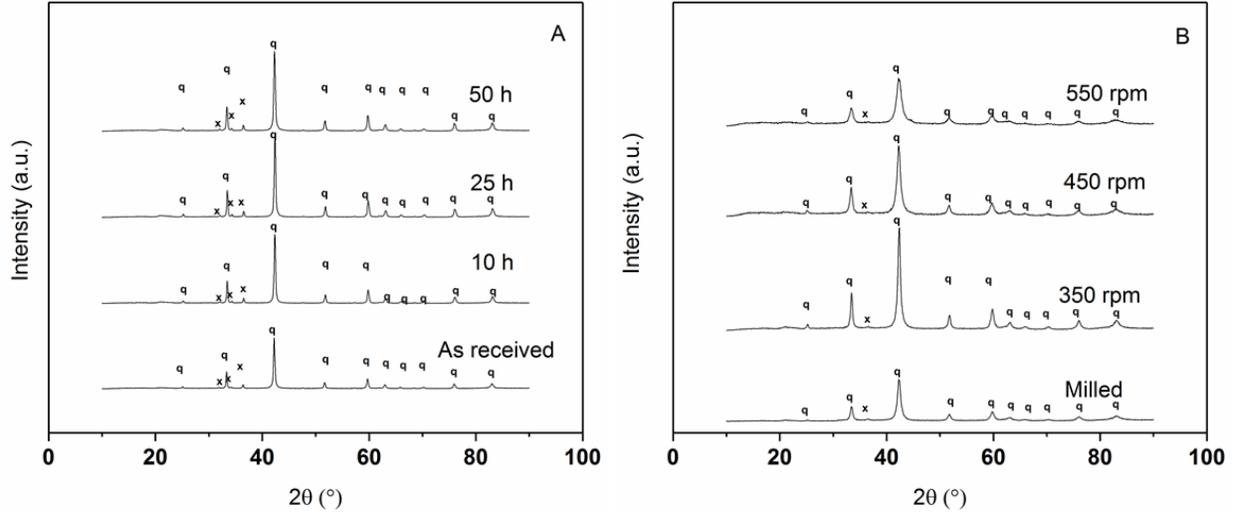

Figure 1: *Ex-situ* PXD patterns of: A) As received, 10 h, 25 h and 50 h; B) Milled, 350 rpm, 450 rpm, and 550 rpm; x = Mg and q = MgB$_2$.

The Rietveld refinement of the As received pattern sets the composition of the starting material to 97 wt% MgB$_2$ and 3 wt% Mg. In these specimens the crystallite size of MgB$_2$ is 830 ± 40 Å and the microstrain is $1.0*10^{-3}$. Compared to the starting material, the value of the crystallite size of the 10 h sample decreases to 630 ± 30 Å. This value remains unchanged also for the pattern of 25 h. The microstrain decreases to $6.5*10^{-4}$ for the pattern of 10 h and to $6.7*10^{-4}$ for the pattern of 25 h. Finally, for the pattern of 50 h the MgB$_2$ crystallite size decreases to 530 ± 25 Å while the microstrain increases to $4.8*10^{-3}$. As expected, the ball milling of the material sensibly altered the microstructural properties of MgB$_2$. In fact, the crystallite size of MgB$_2$ in the Milled pattern is 170 ± 25 Å, and the microstrain equals $2.5*10^{-3}$. These values first increase to 230 ± 10 Å and $1.5*10^{-3}$ for the pattern of 350 rpm and then decrease to 160 ± 10 Å and $2.2*10^{-3}$ for the pattern of 450 rpm and 110 ± 5 Å and $3.6*10^{-3}$ for pattern of 550 rpm.

Table 1: MgB$_2$ crystallite size and microstrain obtained by Rietveld refinement of the PXD patterns reported in Figure 1.

| Sample name | Crystallite size (Å) | Microstrain |
|---|---|---|
| As received | 830 ± 40 | 1.0 *10$^{-3}$ |
| 10 h | 630 ± 30 | 6.5 *10$^{-4}$ |
| 25 h | 630 ± 30 | 6.7 *10$^{-4}$ |
| 50 h | 530 ± 25 | 4.8 *10$^{-4}$ |
| Milled | 170 ± 25 | 2.5 *10$^{-3}$ |
| 350 rpm | 230 ± 10 | 1.5 *10$^{-3}$ |
| 450 rpm | 160 ± 10 | 2.2 *10$^{-3}$ |
| 550 rpm | 110 ± 5 | 3.6 *10$^{-3}$ |

To study the morphological features of the starting materials and the changes which they supposedly underwent during mixing, all the investigated specimens were characterized by the SEM technique, and the results are summarized in Figure 2.

The SEM micrograph of the As received material indicates that most of the MgB$_2$ particles have a size that ranges between 70 and 200 µm whereas the Milled material appears to be constituted mostly from particles of size between 100 and 140 µm. In both cases, smaller fractions of MgB$_2$ particles are observed, *i.e.* between 1 and 10 µm for the As received material and between 30 and 50 µm for the Milled material. The milling of the as-received MgB$_2$ at 550 rpm for increasing time (10 h, 25 h, and 50 h) leaves the average particle size of the main portion of the sample almost unchanged; however, the formation of an increasing portion of particles of size between 1 and 20 µm is observed. The micrographs acquired for the material milled for 50 hours under different rotation regimes (350 rpm, 450 rpm, and 550 rpm) indicate that, in all three cases, the initial average particle size is reduced to 50-70 µm.

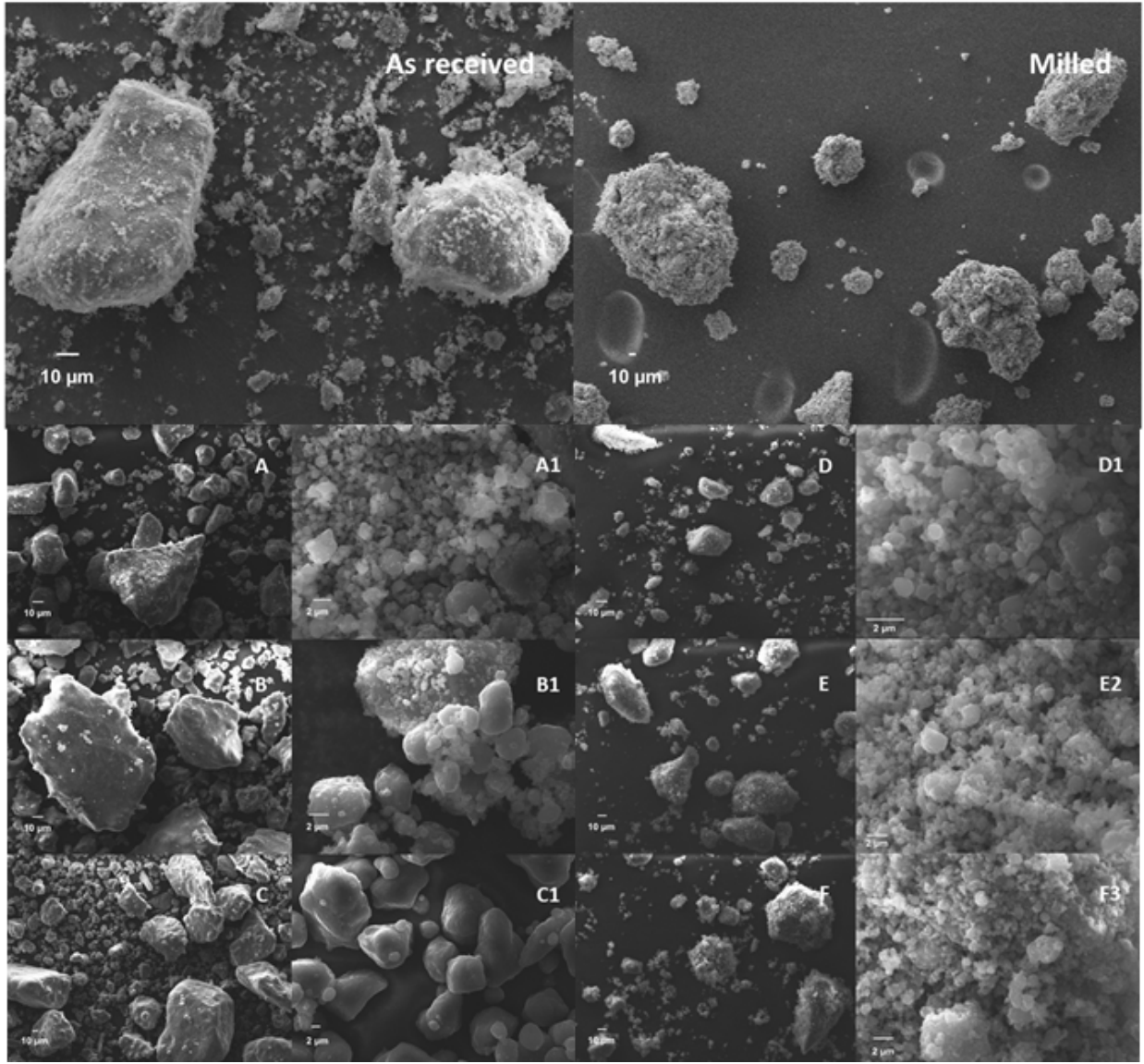

Figure 2: SEM micrograph of: As received, Milled, 10 h (A, A1), 25 h (B, B1), and 50 h (C, C1); 350 rpm (D, D1), 450 rpm (E, E1), and 550 rpm (F, F1).

In order to understand if the applied mechanical treatments led to the hydrogenation of a fraction of the starting $MgB_2$, the obtained specimens were further characterized by means of volumetric technique. The acquired results are summarized in Figure 3. The volumetric curves of the as-received material after milling at 550 rpm for increasing time under hydrogen atmosphere are shown on the left-hand side (A). The volumetric curves obtained for the material milled under a hydrogen atmosphere under different rotation regimes (350 rpm, 450 rpm, and 550 rpm) are shown on the right-hand side (B). These analyses clearly show that, upon milling, the material absorbed quantities of hydrogen which appears to be related to the magnitude of the applied mechanical process. Upon heating, all the samples of Figure 3A start to release hydrogen at about 200 °C and reach a maximum rate of release at about 350 °C. This process appears to be complete shortly after reaching the isothermal period at 450 °C. Overall, the desorption process appears to take place in a single step. The amount of released hydrogen is 0.11 wt%, 0.18 wt% and 0.26 wt% for the 10 h, 25 h and 50 h samples, respectively. The volumetric curves of Figure 3B, similarly to those of Figure 3A, show that the hydrogen release takes place in a single step. The sample 350 rpm starts to release hydrogen at about 250 °C and its dehydrogenation is complete when the temperature reaches 450 °C. The overall amount of released hydrogen is 0.7 wt%. The 450 rpm sample starts releasing hydrogen at about 250 °C until an amount of desorbed hydrogen equal to 1.4 wt% is achieved shortly after reaching the isothermal period at 450 °C. The 550 rpm sample starts releasing hydrogen at about 200 °C and reaches a maximum amount of desorbed hydrogen equal to 2.1 wt% after 100 minutes at 450 °C.

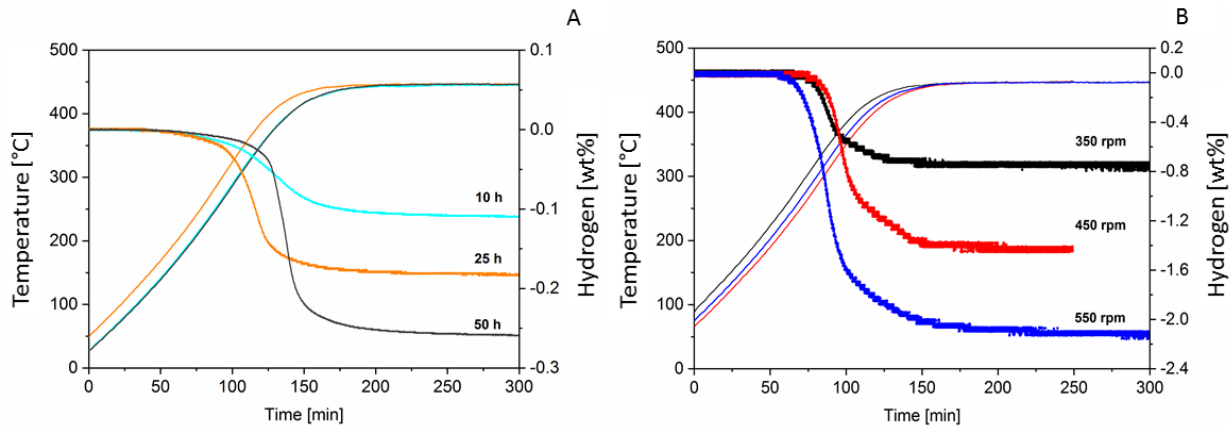

Figure 3: Volumetric analyses of the products of (A) 10 h, 25 h, 50 h, and (B) 350 rpm, 450 rpm, 550 rpm. The measurements were performed at a $p_{H2}$ of 1 bar, heating the material from about 90 °C up to 450 °C with a heating rate of 3 °C/min and then keeping it isothermally at 450 °C for several minutes.

In order to ensure that the observed capacities are due to the release of hydrogen only (boron-containing hydrides might decompose releasing $B_2H_{2)}$, the composition of the gaseous phase released upon heating was investigated by means of an MS device (Figure 4). The monitored gases were $H_2$ and $B_2H_6$. The results of the MS characterizations confirm that the only gas released upon heating is hydrogen. At this point, it is clear that at least a portion of the samples undergoes hydrogenation upon mechanical treatment. Moreover, since no peak belonging to hydrogenated products is visible in the diffraction patterns of Figure 1 the hydride phase/phases must be in an amorphous or nanocrystalline state. Both possibilities are likely. In fact, in our previous work[5] the hydrogenated phases formed *via* ball milling under hydrogen atmosphere were also not visible in the diffraction patterns; moreover, the amorphization process and/or the reduction of the crystallite size to the nano-range size might be also a direct consequence of the hydrogenation process.

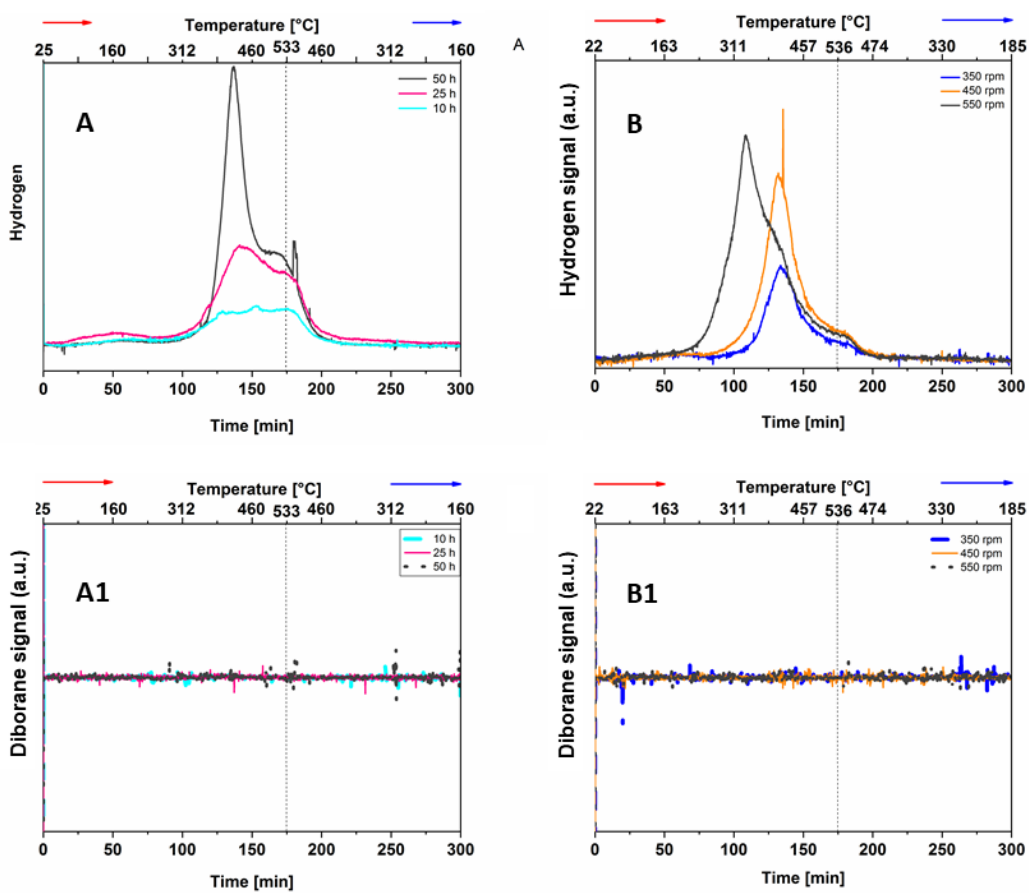

Figure 4: MS signals for $H_2$ (A and B) and $B_2H_6$ (A1 and B1), measured for samples (A, A1; 10 h, 25, 50 h and B, B1; 350 rpm, 450 rpm, 550 rpm).

Aiming to thoroughly characterize the composition of the mechanically treated specimens, the MAS NMR technique was used. Indeed, it is well known that SSNMR provides complementary information with respect to diffraction techniques[49] on several types of materials, independently on their form, either crystalline, nano-/microcrystalline or amorphous.[50] The results of this analysis are summarized in Figure 5. In the $^{11}$B MAS NMR spectra aquired for the 10 h, 25 h, and 50 h samples (Figure 5A) the main resonance of MgB$_2$ at δ($^{11}$B) = 99.3 ppm is visible. It is interesting to notice that in the spectra of 25 h and 50 h a not well-resolved peak starts to be visible at δ($^{11}$B) = -38.4 ppm. This resonance is most likely related to the presence of a small fraction of Mg(BH$_4$)$_2$. Unfortunately, due to the low quality of the Mg(BH$_4$)$_2$ band a quantification of the B-distribution between the phases contained in the sample could not be performed. The spectra reported in Figure 5B clearly show the presence of unreacted MgB$_2$ at δ($^{11}$B) = 99.3 ppm, and compared to the spectra of Figure 5A a better-defined peak at δ($^{11}$B) = -38.4 ppm increases with the increasing milling rotation regime. These results undoubtedly confirm the formation of Mg(BH$_4$)$_2$. Spectra acquired at different spinning speeds, 10 and 12 kHz (see Figure S1 in the Supplementary Information), highlight the presence of an additional peak at 3.5 ppm associated to a new B-containing phase. This signal becomes more pronounced with the increasing of the frequency of rotation in the following order: 350 rpm < 450 rpm < 550 rpm. Since at this moment the nature of this phase is unknown we indicate it as B???. For the spectra displayed in Figure 5B an attempt to calculate the B-distribution between MgB$_2$, Mg(BH$_4$)$_2$, and B??? was carried out. The possibility that some B$_2$O$_3$ might be present in the analyzed samples was considered; however, after a careful analysis of the spectra we established the quantity of this phase to be neglectable.

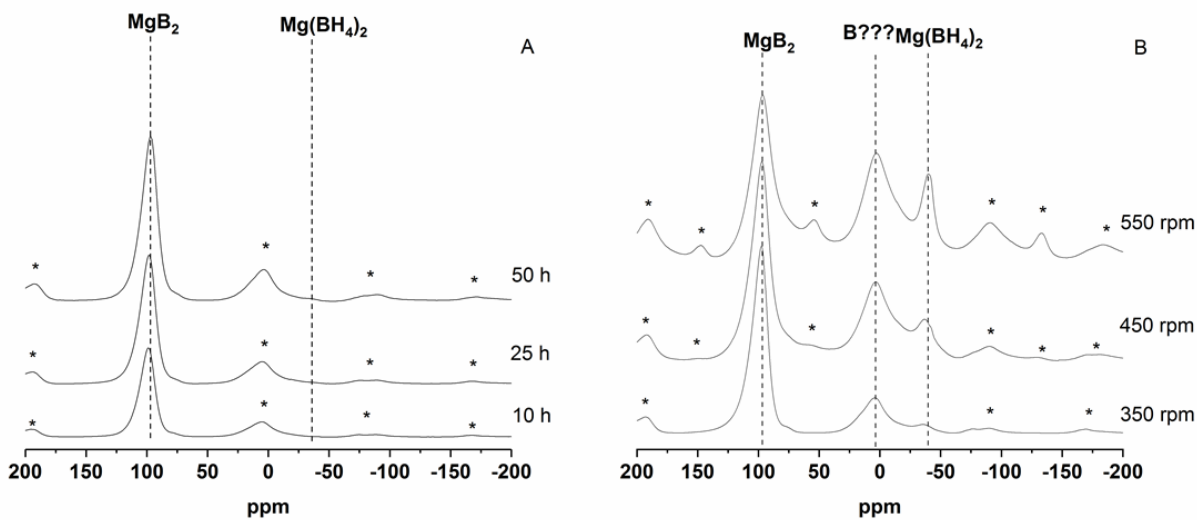

Figure 5: Solid-state $^{11}$B MAS NMR spectra acquired using a spinning speed of 12 KHz for (A) 10 h, 25 h, 50 h, and (B) 350 rpm, 450 rpm, 550 rpm. Spinning sidebands are indicated with an asterisk.

From the integration of the area enclosed in the center and spinning sidebands of the spectra of Figure 5B we estimated the following boron distribution (Table 2):

Table 2: B atoms distribution in $MgB_2$, B and $Mg(BH_4)_2$ for the samples 350 rpm, 450 rpm and 550 rpm.

| Sample | % B atoms in $MgB_2$ | % B atoms in B??? | % B atoms in $Mg(BH_4)_2$ |
|---|---|---|---|
| 350 rpm | 90 | 7.7 | 2.8 |
| 450 rpm | 76.8 | 16.5 | 6.7 |
| 550 rpm | 67.6 | 23.3 | 9.1 |

**Discussion**

The results reported in the previous section indicate that the energy transferred during powder-to-wall collisions in a milling process cannot be neglected. In fact, for the chosen model reaction, the transferred energy not only allowed to refine the material microstructure (Figure 1 and 2) but it also allowed its partial hydrogenation (Figure 3, 4, and 5). It must be noticed that, despite the relatively favorable enthalpy of hydrogenation ($\Delta H_f \sim -40$ KJ/mol$_{H2}$), due to kinetic constrains the hydrogenation of $MgB_2$ under static conditions occurs only under high temperature and high pressure conditions.[37,51,52]

As reported in the literature, a significant number of gas-solid reactions, typically performed under extreme conditions, can be promoted at room temperature *via* mechanochemical route by exploiting ball milling (BM) techniques: reactive BM consists in repeated single impact events under reactive atmospheres, such as hydrogen, carbon monoxide and dioxide, which induce structural and chemical transformations by lowering the activation energy barrier necessary to trigger the reaction.[53,54,55] However, the formation of $MgBH_4$ here described and performed without any grinding media, was quite unexpected. As emerged from the SEM analyses reported in Figure 2 and the volumetric analyses of Figure 3, the particle reactivity seems to be connected to their dimension. For the hydrogenation processes performed under identical rotation regime, extent of time, and consequently milling chamber temperature (lower than 60 °C), the reduced $MgB_2$ particle dimensions allowed achieving a larger degree of conversion of $MgB_2$ into $Mg(BH_4)_2$ *i.e.* 2.1 wt% for 550 rpm and 0.26 wt% for 50 h.  Interestingly, for the material milled for the same extent of time (*e.g.* 350 rpm, 450 rpm, and 550 rpm) a clear correlation between the increased number of rpm and the yield of the conversion is seen (Table 2). Therefore, not only a mechanical input appears to be necessary to initiate the hydrogenation reaction, but also the magnitude of the energy transferred during the collisions appears to play an important role in the conversion of $MgB_2$ into $Mg(BH_4)_2$. To further investigate this aspect, it is plausible to estimate the collision energy for each particle, assuming that their motion during rotation can be approximately treated as the balls in a mechanochemical reactor. For this purpose, it is then possible to exploit the model proposed by Burgio et al.[56] According to this work, the total energy ($\Delta E^*$), released by one particle in a mechanochemical reactor, can be calculated by the equation (5):

$$\Delta E^* = \varphi_b \left\{ -m_b \cdot \left[ \frac{\omega_v^3 \cdot \left(r_v - \frac{d_b}{2}\right)}{\Omega_p} + \Omega_p \cdot \omega_v \cdot R_p \right] \cdot r_v - \frac{d_b}{2} \right\} \quad (5)$$

where $\varphi_b$ is the degree of milling which can be approximated to 1 due to the small diameter of particles, $m_b$ the mass of each particle, *i.e.* $2.32 * 10^{-9}$ kg considering a starting particle dimension of 120 μm, a density of 2.57 g/cm³, and a spherical geometry of the particles, $\Omega_p$ the rotation velocity of the plate (rad/s), $\omega_v$ the rotation velocity of chamber (rad/s), $r_v$ the chamber radius (m), $R_p$ the plate radius (m) and $d_p$ the particle diameter ($1.2 *10^{-4}$ m) . The resulting total energy which each particle dissipated impacting the wall of the chamber is reported in Table 3:

Table 3: Total energy which each particle of 350 rpm, 450 rpm, and 550 rpm dissipates impacting the wall of the chamber.

| Sample name | Total energy per particle |
| --- | --- |
| 350 rpm | 1.39 *$10^{-8}$ J |
| 450 rpm | 2.30 *$10^{-8}$ J |
| 550 rpm | 3.44 *$10^{-8}$ J |

These energy values are 5-6 orders of magnitude smaller than those typically reported for gas-solid reactions activated by mechanochemical input in planetary apparatus.[15,57] On the other hand, it seems enough to impart the threshold energy level to trigger mechanochemical transformation. Starting from this parameter, it is then possible to additionally determine the velocity possessed by each particle during the mixing treatment, which is estimated to be equal to 3.46 m s$^{-1}$, 4.45 m s$^{-1}$, and 5.44 m s$^{-1}$ for 350 rpm, 450 rpm, and 550 rpm, respectively. The particle velocity represents an important parameter that can be exploited for upscaling the gas-solid reaction, for example in a fluidized bed reactor dedicated to the sustainable synthesis, at room temperature, of complex borohydrides. However, despite this first attempt, a deep characterization of the dynamics of the mixing process combined with the determination of a more accurate threshold energy and shear contributions should be carried out.

Further analyzing the $^{11}$B MAS NMR results reported in Figure 5 it appears clear that in all the investigated specimens the milling treatment led to the partial conversion of MgB$_2$ into Mg(BH$_4$)$_2$. However, the additional signal at δ($^{11}$B) = 3.5 ppm revealed the formation of the unidentified phase B???. In order to better understand the nature of this phase and if it contains hydrogen a $^1$H-$^{11}$B CPMAS analysis of the 550 rpm sample was carried out. This analysis is aimed at revealing the presence of $^{11}$B nuclei dipolarly coupled with $^1$H nuclei, *i.e.* $^{11}$B nuclei that present $^1$H nuclei in their chemical environment (close in space). In other words, all signals displayed in the spectra are associated to H-containing phases. The result of this analysis is reported in Figure 6.

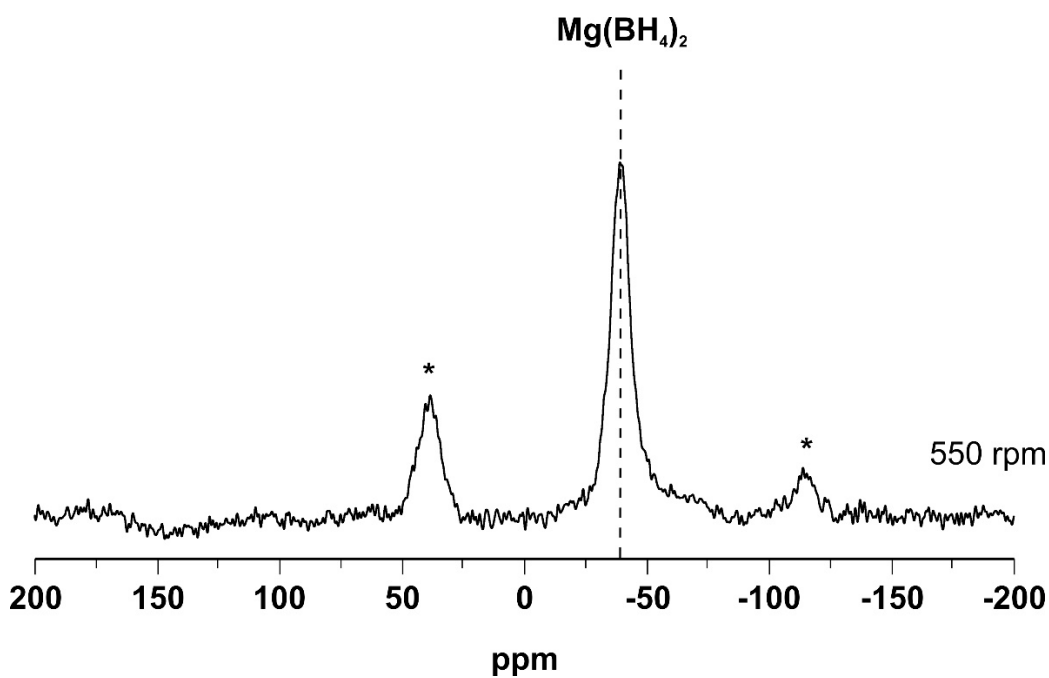

Figure 6: Solid-state $^1$H-$^{11}$B CPMAS NMR spectrum of the 550 rpm sample, acquired using a spinning speed of 10 KHz.

The $^1$H-$^{11}$B CPMAS spectrum of 550 rpm shows only one resonance, at -38.4 ppm, which appeared also in the direct $^{11}$B measurements of Figure 5 and is due to the formation of Mg(BH$_4$)$_2$. Thus, the signal visible in the direct $^{11}$B spectra at 3.5 ppm (Figure 5) corresponds to $^{11}$B domains that do not incorporate hydrogen atoms. Differently from the previously published works on the hydrogenation of MgB$_2$,[5,37] no MgB$_{12}$H$_{12}$ was detected. This might be a direct consequence of the mild energy transferred to the system which prevents the formed Mg(BH$_4$)$_2$ from decomposing following the reaction path reported in equation 3.

To shed further light on the nature of phase B???, first-principles calculation of the NMR shielding tensors was utilized. $^{11}$B NMR parameters for a chosen variety of boron coordination environments were simulated in periodic calculation of bulk systems utilizing density functional theory (DFT). The obtained results (Table 4) indicate that the unidentified peak lies in the scope of boron icosahedral clusters. All optimized structures used for the NMR calculations and their relative coordinates are reported in Figure S2 and Table S1, respectively in the Supplementary Information. Mg-B compounds can be ruled out seeing as their computed $\delta_{iso}$ values lie out of the range of the signal in question, which is localized around 3 ppm. As Table 4 shows, structures containing B$_{12}$ clusters as building units give chemical shifts in the expected range, which is in agreement with work by others.[58,59] The structures of both α- and β-rhombohedral boron phases are closely related and contain B$_{12}$ subunits. Since both allotropes could potentially form under the experimental conditions, they are both likely candidates for the origin of the unidentified signal in the spectrum, although the disordered β-phase is marginally more stable.[60] The lack of diffraction peaks associated to the presence of this boron phase in the PXD analyses of Figure 1 might be justified by the existence of this phase in a nanostructured state.

Table 4: DFT calculated isotropic chemical shifts ($\delta_{iso}$, given in ppm) for symmetry-inequivalent $^{11}$B atoms in selected boron-containing systems.

| Compound | Calculated $\delta_{iso}$, |
|---|---|
| MgB$_2$ | **B-1:** 96.40 (*97.2*)[a] |
| Mg(BH$_4$)$_2$ | **B-1:** -40.0 (*-40.0*)[a] |
| B$_{12}$ icosahedra | **B-1:** 10.54, **B-2:** 7.93 (*5.6*)[b] |
| B$_{12}$ cuboctahedra | **B-1:** 38.50 |
| B$_{36}$ quasi-planar | **B-1:** 59.76, **B-2:** -8.93, **B-3:** 21.99 **B-4:** 27.77, **B-5:** -9.07, **B-6:** -12.19 |
| B$_{28}$ icosahedra | **B-1:** 9.20, **B-2:** 10.70, **B-3:** 4.43, **B-4:** -5.30, **B-5:** 6.83 |
| MgB$_7$ | **B-1:** -31.34, **B-2:** 17.79, **B-3:** 39.88, **B-4:** 1.83, **B-5:** 13.73 |
| MgB$_4$ | **B-1:** -4.06, **B-2:** -51.06 |
| MgB$_3$ | **B-1:** 26.15, **B-2:** -38.75 |

$^a$Experimental chemical shift from this work. $^b$Computed chemical shift from Ref. [58].

When comparing the % of B atoms contained in Mg(BH$_4$)$_2$ (the only B-phase containing hydrogen among the products) in each sample with the amount of hydrogen released during the volumetric measurements (Figure 3) it appears as if the values reported in Table 2 for the % B atoms in Mg(BH$_4$)$_2$ are slightly underestimated. Although we do not possess a piece of direct evidence, this inconsistency might be justified by the presence of nanocrystalline MgH$_2$ among the reaction products. In fact, the NMR analyses of Figures 5 and 6 clearly indicated that during milling an increasing amount of boron is freed into the system due to an incomplete conversion of MgB$_2$ into Mg(BH$_4$)$_2$. Consequently, an increasing amount of Mg should be also present among the reaction products. Considering that in the *ex-situ* PXD analysis of Figure 1 the diffraction peaks of Mg (present already in the as-received material) do not increase in intensity, the additional Mg formed during the milling processes might be in a nanocrystalline state. Thus, it might at least partially react with hydrogen to form MgH$_2$. In fact, it was reported that nanocrystalline Mg can react with hydrogen to form MgH$_2$ already at room temperature.[61]

**Conclusion**

In this work, the effect of the energy transferred through powder-to-wall collisions during the milling process of MgB$_2$ under a hydrogen atmosphere was studied. The obtained pieces of evidence demonstrated that this energy allowed the partial hydrogenation of the starting MgB$_2$ to form Mg(BH$_4$)$_2$. Owing to a series of MAS NMR measurements and numerical simulations it has been possible to find out that consequently to the mechanochemical treatment a fraction of the starting MgB$_2$ is converted in α- or β-rhombohedral boron and Mg/MgH$_2$. Taking advantage of the kinematic model developed by Burgio and co-workers, an attempt to estimate the collision energy for each particle was made for the first time, to the best of the authors' knowledge. The results of this calculation revealed that for the applied measurement conditions and powder features, the energy transferred during a single powder-to-wall impact is 5-6 order of magnitude smaller than the typical energy transferred during ball-to-wall impact during a similar milling process carried out in the presence of grinding media. This work opens a new frontier on the study of the synthesis of material under low-energy mechanochemical input in reactive atmospheres.

**Acknowledgements**

The authors of this work are grateful to Dr. Nils Bergemann and Dr. Pau Nolis for the support in performing part of the presented analyses.

**References**

1. C. Suryanarayana, Progress in Materials Science **46,** 1 (2001).
2. P. Balaz, in *Mechanochemistry in Nanoscience and Minerals Engineering* (Springer, 2008).
3. F. Delogu and G. Cocco, Physical Review B **74,** 035406 (2006).
4. A. Porcheddu, A. Cincotti, and F. Delogu, International Journal of Hydrogen Energy **46,** 967 (2021).
5. C. Pistidda, S. Garroni, F. Dolci, E. G. Bardají, A. Khandelwal, P. Nolis, M. Dornheim, R. Gosalawit, T. Jensen, Y. Cerenius, S. Suriñach, M. D. Baró, W. Lohstroh, and M. Fichtner, Journal of Alloys and Compounds **508,** 212 (2010).
6. J. Barale, S. Deledda, E. M. Dematteis, M. H. Sørby, M. Baricco, and B. C. Hauback, Scientific Reports **10** (2020).
7. I. Z. Hlova, A. Castle, J. F. Goldston, S. Gupta, T. Prost, T. Kobayashi, L. Scott Chumbley, M. Pruski, and V. K. Pecharsky, Journal of Materials Chemistry A **4,** 12188 (2016).
8. E. Grube, C. H. Olesen, D. B. Ravnsbæk, and T. R. Jensen, Dalton Transactions **45,** 8291 (2016).
9. E. Roedern, Y. S. Lee, M. B. Ley, K. Park, Y. W. Cho, J. Skibsted, and T. R. Jensen, Journal of Materials Chemistry A **4,** 8793 (2016).
10. H. Cao, A. Santoru, C. Pistidda, T. M. M. Richter, A. L. Chaudhary, G. Gizer, R. Niewa, P. Chen, T. Klassen, and M. Dornheim, Chemical Communications **52,** 5100 (2016).
11. B. Richter, D. B. Ravnsbæk, N. Tumanov, Y. Filinchuk, and T. R. Jensen, Dalton Transactions **44,** 3988 (2015).
12. Y. J. Tan, Z. Zhang, F. J. Wang, H. H. Wu, and Q. H. Li, RSC Advances **4,** 35635 (2014).
13. D. B. Ravnsbæk, E. A. Nickels, R. Černý, C. H. Olesen, W. I. F. David, P. P. Edwards, Y. Filinchuk, and T. R. Jensen, Inorganic Chemistry **52,** 10877 (2013).
14. M. Polanski, T. K. Nielsen, I. Kunce, M. Norek, T. Płociński, L. R. Jaroszewicz, C. Gundlach, T. R. Jensen, and J. Bystrzycki, International Journal of Hydrogen Energy **38,** 4003 (2013).
15. T. T. Le, C. Pistidda, J. Puszkiel, C. Milanese, S. Garroni, T. Emmler, G. Capurso, G. Gizer, T. Klassen, and M. Dornheim, Metals **9** (2019).

| 16 | R. Hardian, C. Pistidda, A. L. Chaudhary, G. Capurso, G. Gizer, H. Cao, C. Milanese, A. Girella, A. Santoru, D. Yigit, H. Dieringa, K. U. Kainer, T. Klassen, and M. Dornheim, International Journal of Hydrogen Energy **43,** 16738 (2018). |
| 17 | C. Pistidda, N. Bergemann, J. Wurr, A. Rzeszutek, K. T. Møller, B. R. S. Hansen, S. Garroni, C. Horstmann, C. Milanese, A. Girella, O. Metz, K. Taube, T. R. Jensen, D. Thomas, H. P. Liermann, T. Klassen, and M. Dornheim, Journal of Power Sources **270,** 554 (2014). |
| 18 | S. Garroni, C. Pistidda, M. Brunelli, G. B. M. Vaughan, S. Suriñach, and M. D. Baró, Scripta Materialia **60,** 1129 (2009). |
| 19 | R. Černý, D. B. Ravnsbæk, P. Schouwink, Y. Filinchuk, N. Penin, J. Teyssier, L. Smrčok, and T. R. Jensen, Journal of Physical Chemistry C **116,** 1563 (2012). |
| 20 | L. H. Rude, M. Corno, P. Ugliengo, M. Baricco, Y. S. Lee, Y. W. Cho, F. Besenbacher, J. Overgaard, and T. R. Jensen, Journal of Physical Chemistry C **116,** 20239 (2012). |
| 21 | R. Černý, D. B. Ravnsbæk, G. Severa, Y. Filinchuk, V. D'Anna, H. Hagemann, D. Haase, J. Skibsted, C. M. Jensen, and T. R. Jensen, Journal of Physical Chemistry C **114,** 19540 (2010). |
| 22 | X. Z. Xiao, L. X. Chen, X. L. Fan, X. H. Wang, C. P. Chen, Y. Q. Lei, and Q. D. Wang, Applied Physics Letters **94,** 041907 (2009). |
| 23 | L. Li, Z.-C. Zhang, Y.-J. Wang, L.-F. Jiao, and H.-T. Yuan, Rare Metals **36,** 517 (2017). |
| 24 | J. Huot, D. B. Ravnsbæk, J. Zhang, F. Cuevas, M. Latroche, and T. R. Jensen, Progress in Materials Science **58,** 30 (2013). |
| 25 | H. I. Schlesinger, Journal of the American Chemical Society **62,** 3421 (1940). |
| 26 | H. I. Schlesinger, Journal of the American Chemical Society **13,** 1765 (1941). |
| 27 | H. I. Schlesinger and H. C. Brown, Journal of the American Chemical Society **62,** 3429 (1940). |
| 28 | E. Wiberg and W. Henle, Zeitschrift fur Naturforschung - Section B Journal of Chemical Sciences **7,** 582 (1952). |
| 29 | E. Wiberg and W. Henle, Zeitschrift fur Naturforschung - Section B Journal of Chemical Sciences **7,** 579 (1952). |
| 30 | H. I. Schlesinger, H. C. Brown, A. E. Finholt, J. R. Gilbreath, H. R. Hoekstra, and E. K. Hyde, Journal of the American Chemical Society **75,** 215 (1953). |
| 31 | H. I. Schlesinger, H. C. Brown, B. Abraham, A. C. Bond, N. Davidson, A. E. Finholt, J. R. Gilbreath, H. Hoekstra, L. Horvitz, E. K. Hyde, J. J. Katz, J. Knight, R. A. Lad, D. L. Mayfield, L. Rapp, D. M. Ritter, A. M. Schwartz, I. Sheft, L. D. Tuck, and A. O. Walker, Journal of the American Chemical Society **75,** 186 (1953). |
| 32 | T. Matsunaga, F. Buchter, P. Mauron, M. Bielman, Y. Nakamori, S. Orimo, N. Ohba, K. Miwa, S. Towata, and A. Züttel, Journal of Alloys and Compounds **459,** 583 (2008). |
| 33 | K. Chlopek, C. Frommen, A. Leon, O. Zabara, and M. Fichtner, Journal of Materials Chemistry **17,** 3496 (2007). |
| 34 | V. A. Kuznetsov and T. N. Dymova, Russian Chemical Bulletin **20** (1971). |
| 35 | G. L. Soloveichik, Y. Gao, J. Rijssenbeek, M. Andrus, S. Kniajanski, R. C. Bowman Jr, S. J. Hwang, and J. C. Zhao, International Journal of Hydrogen Energy **34,** 916 (2009). |
| 36 | D. S. Stasinevich and G. A. Egorenko, Russ. J. Inorg. Chem. **13,** 341 (1968). |
| 37 | G. Severa, E. Rönnebro, and C. M. Jensen, Chemical Communications **46,** 421 (2010). |
| 38 | P. Zanella, L. Crociani, N. Masciocchi, and G. Giunchi, Inorganic Chemistry **46,** 9039 (2007). |
| 39 | R. Černý, Y. Filinchuk, H. Hagemann, and K. Yvon, Angewandte Chemie - International Edition **46,** 5765 (2007). |
| 40 | Y. Filinchuk, R. Cerny, and H. Hagemann, Chemistry of Materials **21,** 925 (2009). |
| 41 | J. H. Her, P. W. Stephens, Y. Gao, G. L. Soloveichik, J. Rijssenbeek, M. Andrus, and J. C. Zhao, Acta Crystallographica Section B: Structural Science **63,** 561 (2007). |

42. Y. Filinchuk, B. Richter, T. R. Jensen, V. Dmitriev, D. Chernyshov, and H. Hagemann, Angewandte Chemie - International Edition **50,** 11162 (2011).
43. K. Persson and M. Project, (2020).
44. O. Zavorotynska, A. El-Kharbachi, S. Deledda, and B. C. Hauback, International Journal of Hydrogen Energy **41,** 14387 (2016).
45. L. Lutterotti, Nuclear Instruments and Methods in Physics Research Section B: Beam Interactions with Materials and Atoms **268,** 334 (2010).
46. C. J. Pickard and F. Mauri, Physical Review B **63,** 245101 (2001).
47. G. Kresse and J. Furthmüller, Computational Materials Science **6,** 15 (1996).
48. J. P. Perdew, K. Burke, and M. Ernzerhof, Physical Review Letters **77,** 3865 (1996).
49. A. Rossin, M. R. Chierotti, G. Giambastiani, R. Gobetto, and M. Peruzzini, CrystEngComm **14,** 4454 (2012).
50. J. Sanz, in *Defects and Disorder in Crystalline and Amorphous Solids*, p. 157.
51. Y. S. Liu, L. E. Klebanoff, P. Wijeratne, D. F. Cowgill, V. Stavila, T. W. Heo, S. Kang, A. A. Baker, J. R. I. Lee, T. M. Mattox, K. G. Ray, J. D. Sugar, and B. C. Wood, International Journal of Hydrogen Energy **44,** 31239 (2019).
52. C. Sugai, S. Kim, G. Severa, J. L. White, N. Leick, M. B. Martinez, T. Gennett, V. Stavila, and C. Jensen, ChemPhysChem **20,** 1301 (2019).
53. P. Baláž, M. Achimovičová, M. Baláž, P. Billik, Z. Cherkezova-Zheleva, J. M. Criado, F. Delogu, E. Dutková, E. Gaffet, F. J. Gotor, R. Kumar, I. Mitov, T. Rojac, M. Senna, A. Streletskii, and K. Wieczorek-Ciurowa, Chemical Society Reviews **42,** 7571 (2013).
54. F. Delogu, G. Mulas, and S. Garroni, Applied Catalysis A: General **366,** 201 (2009).
55. F. Torre, V. Farina, A. Taras, C. Pistidda, A. Santoru, J. Bednarcik, G. Mulas, S. Enzo, and S. Garroni, Powder Technology **364,** 915 (2020).
56. N. Burgio, A. Iasonna, M. Magini, S. Martelli, and F. Padella, Il Nuovo Cimento D **13,** 459 (1991).
57. S. Garroni, C. B. Minella, D. Pottmaier, C. Pistidda, C. Milanese, A. Marini, S. Enzo, G. Mulas, M. Dornheim, M. Baricco, O. Gutfleisch, S. Suriñach, and M. D. Baró, International Journal of Hydrogen Energy **38,** 2363 (2013).
58. M. Ludwig and H. Hillebrecht, Physical Chemistry Chemical Physics (2021).
59. C. L. Turner, R. E. Taylor, and R. B. Kaner, The Journal of Physical Chemistry C **119,** 13807 (2015).
60. T. Ogitsu, E. Schwegler, and G. Galli, Chemical Reviews **113,** 3425 (2013).
61. J. Lu, Y. J. Choi, Z. Z. Fang, H. Y. Sohn, and E. Rönnebro, Journal of the American Chemical Society **132,** 6616 (2010).